# A METHOD TO ESTABLISH GPS GRID IONOSPHERIC CORRECT MODEL

Shuanggen Jin

*Shanghai Astronomical Observatory, Chinese Academy of Sciences, Shanghai 200030, China.*
E-mail: sgjin@shao.ac.cn or sg.jin@unsw.edu.au

**Abstract:** The ionospheric influence is one of the largest error sources in GPS positioning and navigation after closing the Selective Availability (SA). Therefore, it is available to establish a real time ionospheric correction model to eliminate or mitigate the ionospheric influence. In this paper, a new method, Hardy Function Interpolation method, is presented to establish a high precision grid ionospheric model (GIM) over the Yangtze River Delta using continuous GPS data of SIGAN network, and the internal and external accuracy of TEC from the GIM are evaluated. It has shown that the real time and high precision GIM over the Yangtze River delta is well established using the Hardy Function Interpolation. The internal and external accuracy of TEC from the GIM are all smaller than 0.3m and better than the methods of distance weight function of WAAS and spherical harmonic function. In addition, these methods are further used to initially investigate and analyze the seasonal variations of TEC over Yangtze River delta.

## INTRODUCTION

The ionospheric affects the transmission of electromagnetic waves, which can result in disturbance or intermission of signals (Jin and Zhu, 2002). The magnitude of this effect is determined by the amount of total electron content (TEC) and the frequency of electromagnetic waves. The Global Positioning System (GPS) is a satellite microwave technique whose signals are transmitted on microwave (L-band) carriers through the Earth's atmosphere, and it inevitably suffers the ionospheric effect. The magnitude of influence on GPS signals is usually in the range from a few meters to tens of meters but it could reach more than 100 meters during severe ionosphere storms (e.g. Liu et al, 1999). Therefore, the ionospheric effect must be estimated so that a correction can be made to eliminate or mitigate for high precision GPS positioning applications. Additionally, precise estimates of ionospheric effect are also important for space weather researches and predictions of the ionospheric events.

Now GPS receivers of double-frequency can monitor the variation of ionospheric electron content (e.g. Yuan and Ou, 2001; Zhao X F, 2003). Establishing a high precision GIM is one of the critical steps for investigating the ionosphere using GPS data. IGS Ionosphere Working Group also uses the GPS-based GIM to research and investigate the variation and action of global ionosphere. The ionosphere is related to the time, season, location, and activity of the Sun, and therefore, the difference of ionospheric effects on the propagation of radio in different local areas or time is very large (e.g. Yuan Y and Ou J, 2001). For some (grid) ionospheric models established by GPS, the parameters of the ionospheric models are usually determined by fitting all GPS data in the large area, which limits the precision of the model due to neglecting the local character of ionosphere and also is not suitable for analyzing the local effects of ionospheric parameters (e.g. Yuan and Ou, 2001; Liu et al, 1999; Zhao X F, 2003). For a small and local region, many methods are used to investigate grid ionospheric model using GPS data from ground GPS networks in the past several years, such as the methods of distance weight function of WAAS (e.g. Skone, S., 1998), spherical harmonic function and the polynomial function (e.g. Schaer, 1999; Komjathy, 1997). In this paper, a new method, Hardy Function Interpolation method, is presented to establish a high precision grid ionospheric correction model over the Yangtze River delta region using GPS data of the Shanghai Integration GPS Application Network (SIGAN). The internal and external accuracy of TEC from the GIM are evaluated to investigate the possibility that the GIM provides high precision ionospheric correction.

## HARDY FUNCTION INTERPOLATION

The current empirical ionospheric correction models, such as the Klobuchar model, could only correct about 50% of the total ionosphere effects (e.g. Klobuchar, 1987). Therefore, the more precise ionosphere model is required. As an approximation, the ionosphere may be considered to be a thin layer, i.e. an ionospheric spherical shell, to a height of 350 km above the earth's surface (e.g. Otsuka Y, 2002). The grid ionospheric model values are the vertical ionospheric delays or vertical total ionospheric content (VTEC) at the specified Ionospheric Grid Points (IGPs. i.e. the intersection points of the selected longitude and latitude lines) covering the area.

In 1977, Hardy developed the Hardy Function Interpolation (HFI) method and applied it to analyze crustal vertical deformation (e.g. Liu et al, 2001). We here apply the HFI method to establish GPS grid ionospheric model over a small local region. For the GPS site coordinate ($B_i$, $L_j$), the VTEC of fitting model can be written as (e.g. Liu et al, 2001):

$$\text{VTEC}(B, L) = \sum_{i=1}^{n} a_i Q(B, L, B_i, L_i) \qquad (1)$$

where VTEC(B, L) is the VTEC of the measurement points in a single plane, ($B_i$ and $L_i$) is the coordinate of grid node, and $a_i$ is the vector of the VTEC value at the grid point. The corresponding core function can be expressed as follows:

$$Q(B, L, B_i, L_i) = [(B - B_i)^2 + (L - L_i)^2 + \varepsilon^2]^\beta \qquad (2)$$

In Eq. (2), $\varepsilon^2$ and $\beta$ are the smooth factor, whose empirical values are 0.01 and 0.5 respectively (Liu et al, 2001). If there are m measurement points ($B_j$, $L_j$) and n grid nodes ($B_i$, $L_i$), and let n grid nodes ($B_i$, $L_i$) as central sites of core function $Q$. Thus, the VTEC of each GPS epoch in the Eq. (1) can be written as follows:

$$v_{(VTEC)} = Qa \qquad (3)$$

Namely:

$$\begin{bmatrix} v_1(B_1, L_1)_{VTEC} \\ v_1(B_2, L_2)_{VTEC} \\ \vdots \\ v_1(B_m, L_m)_{VTEC} \end{bmatrix} = \begin{bmatrix} Q(B_1, L_1, B_{i1}, L_{i1}) & \cdots & Q(B_1, L_1, B_{in}, L_{in}) \\ \vdots & & \vdots \\ Q(B_m, L_m, B_{i1}, L_{i1}) & \cdots & Q(B_m, L_m, B_{in}, L_{in}) \end{bmatrix} \begin{bmatrix} a_1 \\ a_2 \\ \vdots \\ a_n \end{bmatrix} \qquad (4)$$

The error equation is as follows:

$$\Delta v = Qa - v_{VTEC} \quad (5)$$

Through the Eq. (4) we can obtain the vector *a* by a weighted least squares adjustment to all GPS observations in each epoch, namely

$$a = (Q^T P_v Q)^{-1} Q^T P_v v \quad (6)$$

where $P_v$ is the weight matrix of the VTEC. And that, we can obtain the VTEC of random grid sites, namely: $\text{VTEC}_h = Q_h^t a$, where $Q_h^t = (Q_{h1} \quad Q_{h2} \quad \cdots \quad Q_{hi})$.

## SIGAN network

In order to accurately monitor and predict the atmospheric state parameters over the Yangtze River delta region, China, starting from 1999 SHAO (Shanghai Astronomical Observatory, Chinese Academy of Sciences), Shanghai meteorology bureau, and Shanghai institute of surveying and mapping jointly established the Shanghai Integrated GPS Application Network (SIGAN), together with the international GPS service (IGS) station Shanghai (SHAO). The SIGAN network is a 24-hour continuous operating geodetic GPS arrays consisting 14 GPS stations in which the dual frequency Ashtech$^{TM}$ GPS receivers were installed, covering the Yangtze River delta region (Figure 1.). Since July 2002, the SIGAN network has been operating normally. All the sites transmit GPS data of 30 sec and surface meteorological data of 6 min to the center of data process (SHAO) every 30 min.

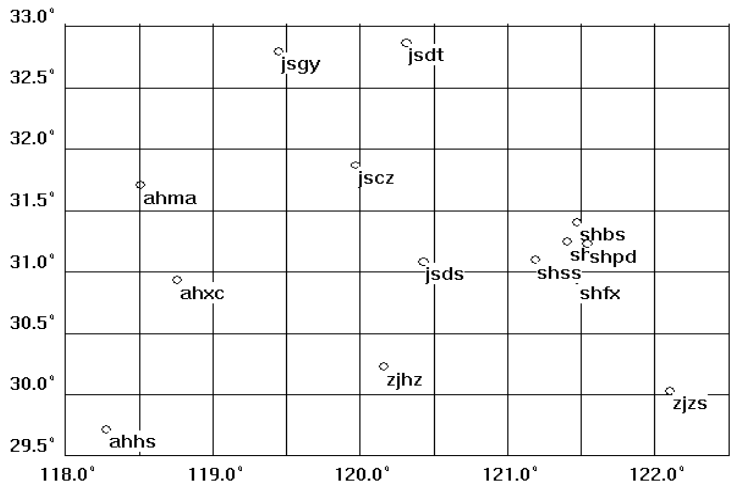

Figure 1. GPS site distribution of Shanghai Integration GPS application Network

## RESULTS AND APPLICATIONS

To measure the TEC over the Yangtze River delta region, we have developed a Grid Ionospheric model of TEC over in the Yangtze River Delta using the Hardy Function Interpolation method with a spatial resolution of $0.2^o \times 0.2^o$. The TEC of each grid site can be mapped for every 30 seconds from the GPS dual frequency phase and code observable. For example, at 6:00(UT) on September 7, 2003, the distribution of VTEC of each grid sites with a spatial resolution of $0.2^o \times 0.2^o$ and contour of VTEC over the Yangtze River delta region are shown clearly in Figure 2 ($1\text{TECU} = 1 \times 10^{16}\ e/m^2$), and the mean standard deviation of the TEC estimated from this GIM is less than 1 TECU. The VTEC variations every six hours over the Yangtze River region are shown in Figure 3.

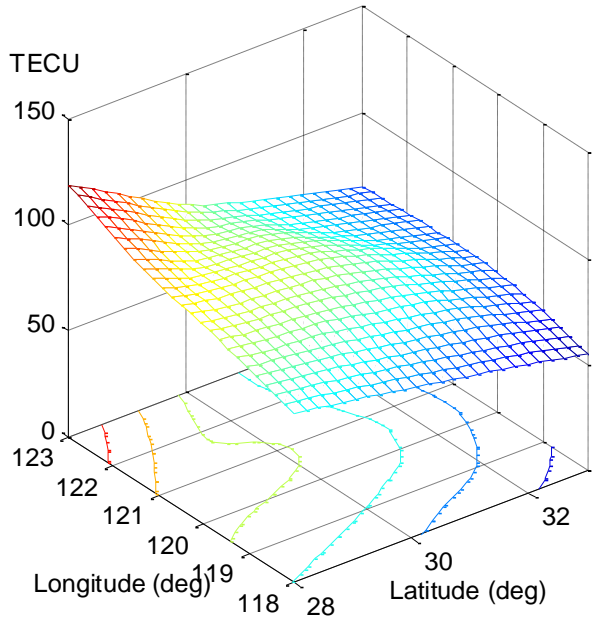

Figure 2. The VTEC and contour over the Yangtze River delta region (UT: 6:00)

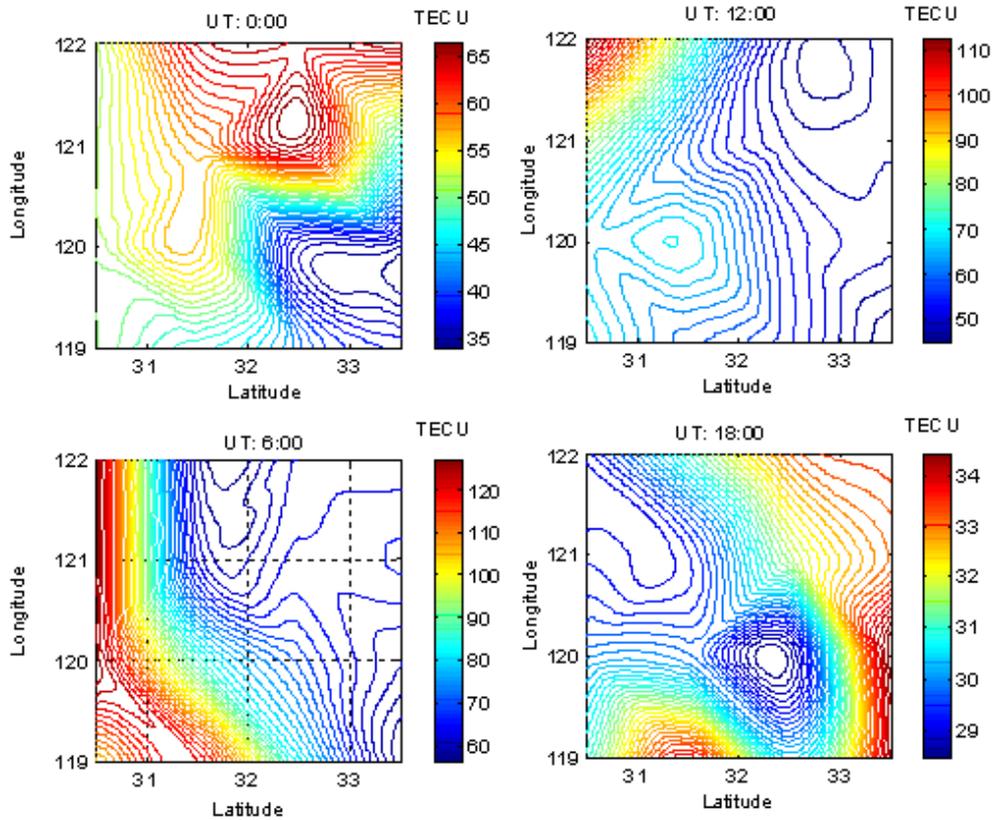

Figure 3. VTEC variations over the Yangtze River delta region on September 7, 2003

The internal and external accuracy of VTEC reflect the quality of VTEC estimated from the Grid Ionospheric Model, where the internal accuracy denotes the root mean square (RMS) of the discrepancy between the observing and fitting VTEC value, and the external accuracy represents the root mean square (RMS) of VTEC. In order to test the reliability of VTEC estimated from the Grid Ionospheric Model of HFI (Hardy Function Interpolation), we develop another two kinds of Grid Ionospheric Models using the distance weight function of Wide Area Augment System (WAAS) (e.g. Chao and Tsai 1996; Skone S, 1998; Liu et al, 1999) and spherical harmonic function method (Schaer, 1999), respectively. The corresponding internal and external accuracy of VTEC are derived respectively. Figure 4 shows comparisons of the internal and external accuracy of VTEC from three Grid Ionospheric Models. It has seen that the accuracies from the three methods are almost consistent, and the mean internal accuracy is 0.1 m, and the mean external accuracy is 0.3m. However, the internal and external accuracy from Hardy Function Interpolation are slightly better than the ones from other two methods. Therefore, it is available to establish the GIM with the HFI method.

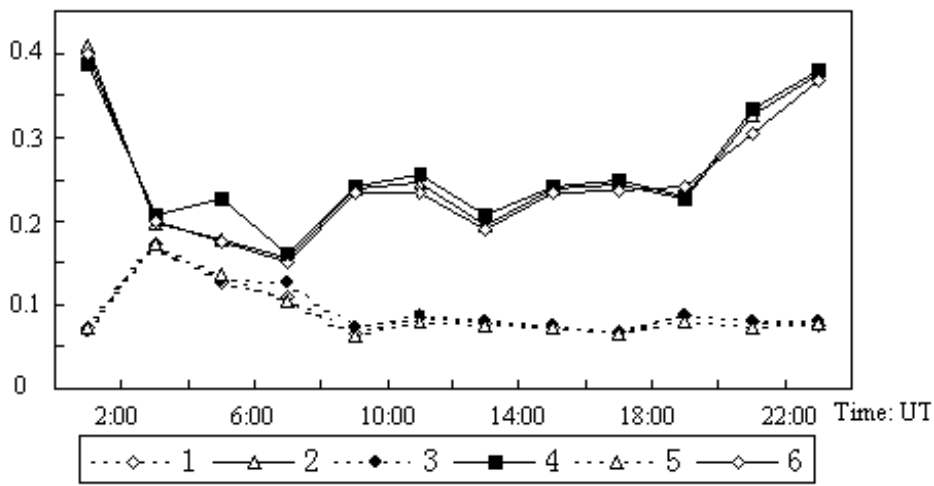

Figure 4. Comparisons of the internal and external accuracy of TEC from GIMs

(Note: the vertical axis is the accuracy of vertical delay in L1 frequency (unit: meter); 1, 3 and 5 represent the internal accuracy from Hardy Function Interpolation, distance weight function of WAAS and spherical harmonic function methods respectively; and 2, 4 and 6 represent the external accuracy from Hardy Function Interpolation, distance weight function of WAAS and spherical harmonic function methods respectively)

In addition, the gridded TECs over the Yangtze River Delta in January (winter), April (spring), July (summer) and October (autumn) 2003 are further derived from the SIGAN network using the Hardy Function Interpolation, distance weight function of WAAS and spherical harmonic function methods. These results are almost close to each other with less than 5 TECU. The seasonal variations of TEC over the Yangtze River delta are investigated and analyzed. The greater TECs are found in the equinoxes, i.e. the so-called semiannual anomaly. Generally speaking, the TEC is higher in winter than in summer from in daytime, but opposite in nighttime. This is due mainly to the fact that the loss rate of electron density depends mainly on the molecular nitrogen concentration [N2] with some contribution from molecular oxygen concentration [O2] in F region, while the production rate depends on the atomic oxygen concentration [O]. The composition changes can be a result of the equinox anomaly effect of TEC due to the convection of atomic oxygen from the summer to the winter hemisphere (Torr and Torr, 1973). In addition, Torr and Torr (1973) also suggested that the semiannual TEC anomaly is due to the semiannual variations in neutral densities associated with geomagnetic and auroral activity. Millward et al. (1996), using the coupled thermosphere –ionosphere–plasmasphere model (CTIP), showed that the offset of the geomagnetic axis from Earth's spin axis is the cause of the semiannual anomaly of noontime NmF2 in the South American sector.

## Conclusion and discussion

The Hardy Function Interpolation (HFI), distance weight function of WAAS and spherical harmonic function methods are used to establish the grid ionospheric models (GIM) over the Yangtze River Delta using continuous GPS data of SIGAN network. The internal and external accuracy of VTEC from three methods are all better than 0.3m, but the internal and external accuracy from the HFI are slightly better than ones from other two methods. Therefore it is reliable to establish a real time and high precision GIM over the Yangtze River delta region with the Hardy Function Interpolation method. Using the regional GIM model from the Hardy Function Interpolation it can produce more detailed instantaneous maps of the regional ionosphere, as compared to other global models, e.g., monitoring and forecasting the local ionospheric events over the Yangtze River delta region.

Furthermore, the seasonal variations of gridded TEC over the Yangtze River Delta in 2003 are initially investigated and analyzed from the SIGAN network using the Hardy Function Interpolation, distance weight function of WAAS and spherical harmonic function methods. It has shown that the greater TECs are found in the equinoxes, while the TEC is higher in winter than in summer from in daytime, but opposite in nighttime. This is due mainly to the composition changes in the molecular nitrogen concentration [N2], molecular oxygen concentration [O2] and atomic oxygen concentration [O] in F region. In the future, more detailed analysis in spatio-temporal variation characteristics of TEC over the Yangtze River Delta will be analyzed using multi-satellite data (e.g. GPS and TOPEX data).

## APPENDIX 1.REFERENCES